\documentclass[review]{elsarticle}
\usepackage[utf8]{inputenc}

\usepackage{amsmath}
\usepackage{amssymb}
\usepackage{subfig}
\usepackage{amsfonts}
\usepackage[ruled,vlined,linesnumbered]{algorithm2e}
\usepackage{algpseudocode}
\usepackage{graphicx}
\usepackage{textcomp}
\usepackage{soul}
\usepackage{xcolor}
\usepackage{adjustbox, booktabs, tabularx, makecell, array}
\usepackage{float}

\DeclareMathOperator*{\argmin}{argmin} 

\title{Towards informed partitioning for load balancing: a proof-of-concept}

\author[1]{Anthony Boulmier\corref{cor1}}
\author[2]{Nabil Abdennadher}
\author[1]{Bastien Chopard}

\address[1]{University of Geneva, Departement of Computer Science, Route de Drize 7, 1227 Carouge, Switzerland
}
\address[2]{University of Applied Sciences and Arts, Western Switzerland (HES-SO), Rue de la Prairie 4, 1202 Geneva, Switzerland}

\cortext[cor1]{Corresponding author\\Email address: anthony.boulmier@unige.ch}

\begin{document}

\begin{frontmatter}

\begin{abstract}

Most parallel applications suffer from load imbalance, a crucial performance degradation factor. In particle simulations, this is mainly due to the migration of particles between processing elements, which eventually gather unevenly and create workload imbalance. Dynamic load balancing is used at various iterations to mitigate load imbalance, employing a partitioning method to divide the computational space evenly while minimizing communications. In this paper, we propose a novel partitioning methodology called ``informed partitioning''. It uses information based on the evolution of the computation to reduce the load balancing growth and the number of load balancing calls. In this paper, we illustrate informed partitioning by proposing a new geometric partitioning technique for particles simulations. This technique is derived from the well-known recursive coordinate bisection and employs the velocity of the particles to guide the bisection axis. To properly compare the performance of our new method with existing partitioning techniques during application execution, we introduce an effort metric based on a theoretical model of load balanced parallel application time. We propose a proof-of-concept of informed partitioning, through a numerical study, on three N-Body simulations with various particle dynamics, and we discuss its performance against popular geometric partitioning techniques. Moreover, we show that our effort metric can be used to rank partitioning techniques by their efficiency at any time point during the simulation. Eventually, this could be used to choose the best partitioning on the fly. In the numerical study, we report that our novel concept increases the performance of two experiments out of three by up to $76\%$ and $15\%$, while being marginally slower by only $3\%$ in one experiment. Also, we discuss the limitations of our implementation of informed partitioning and our effort metric.

\end{abstract}


\end{frontmatter}

\section{Introduction}

In most parallel applications, performance is crucial to solving large and complex problems. One of the most critical performance degradation factors is load imbalance. Usually, load imbalance is mitigated through dynamic load balancing mechanisms. These mechanisms divide the computational elements into several pieces (partitioning algorithm) and distribute them to processing elements (mapping algorithm), such that the workload is evenly distributed and the communications are minimized. Essentially, load balancing consists of finding an approximate solution to a balanced graph partitioning problem, which is known to be NP-Complete~\cite{Garey1979}. Moreover, this raises two challenging questions: \textit{how to load balance?} (i.e., which partitioning technique should I use?) and \textit{when to load balance?} (i.e., at which iteration should I trigger the load balancing mechanism?)~\cite{Pearce2014} to obtain the maximal performance.

Over time, many partitioning algorithms that provide good solutions to the load balancing problem have been developed. Among them, the most famous are recursive coordinate bisection (RCB)~\cite{Simon1997HowBisection}, space-filling curves (SFC)~\cite{Pilkington1996DynamicCurves}, recursive spectral bisection~\cite{VanDriessche1995AnBalancing}, and METIS (multilevel k-way)~\cite{Karypis1998AGraphs}. More recently, novel geometric partitioning techniques have been proposed.
Zhakhovskii~et~al.~\cite{Zhakhovskii2005ASimulation}, Fattebert~et~al.~\cite{Fattebert2012}, and later Egorova~et~al.~\cite{Egorova2019ParallelSubdomains} introduced partitioning methods based on Voronoï tesselations for molecular dynamic applications (particle simulations). In particular, Zhakhovskii~et~al.~\cite{Zhakhovskii2005ASimulation} load balances at regular intervals by replacing the Voronoi site according to the local mass center combined with that of the Voronoï neighbors that move, while Fattebert~et~al.~\cite{Fattebert2012} employ a gradient method and an estimation of the work per volume to adjust the processing elements' workload by moving Voronoï sites. Note that Fattebert~et~al.~\cite{Fattebert2012} were the first to propose such a partitioning method in 3D. Finally, Egorova~et~al.~\cite{Egorova2019ParallelSubdomains} consider Voronoï sites as weighted bodies, where the weight depends on the processing element's load. They compute the displacement required by a Voronoï site to be balanced with respect to the neighboring domains using multi-body terms. Note that their algorithm requires many iterations to converge to a balanced state. 
Begau~et~al.~\cite{Begau2015} proposed to decompose the computational domain into cubical subdomains of equal volume. Each subdomain is assigned to one processing element that computes the particles spatially located in their cube. The load is balanced using a diffusion process that moves each vertex of every subdomain independently to achieve load balancing. All processing elements that share the current vertex (i.e., $8$ vertices per cuboid, each of them is shared by $8$ neighboring processing elements in 3D and $4$ in 2D) cooperate to move it in the direction of the center of mass of the particles belonging to the subdomains of the neighboring processing elements. In 2D, the process remains the same, but the subdomains are squares. 
Deveci~et~al.~\cite{Deveci2016} developed an improved version of the RCB algorithm (available in Zoltan2~\cite{Boman2012Zoltan2}) that performs recursive multisection along the axis of the largest dimension. Moreover, data are migrated while partitioning. These two improvements allow the minimization of data movement compared to the classical implementation of RCB. 

Recently, Hirschmann~et~al.~\cite{Hirschmann2016} studied the behavior of standard partitioning techniques in particle simulation using a load dynamic metric. This metric uses the norm of the difference between the load distribution at two iterations and shows that load balancing techniques are not equal with respect to the load dynamic (i.e., load imbalance growth). They define the iteration's load distribution as the empirical probability density function that estimates the probability to observe a processing element with a given load at a particular iteration.  Furthermore, in a previous work~\cite{Boulmier2021OptimalCriteria}, we confirmed the observations about the load dynamic made by Hirschmann~et~al. These works indicate that load balancing algorithms induce an ``effort'' on the application. This effort comprises the load balancing cost, the imbalance correcting capability, and the load imbalance growth between two consecutive load balancing calls. Hence, a good partitioning should not only balance the load at the time it is applied, but it should also have a long lifetime that is resilient to the growth of a new imbalance and have a small CPU time cost.  Unfortunately, we identified that state-of-the-art partitioning techniques, such as the ones above, lack considering the imbalance growth and, in particular, adapting to it using the data at disposal within each ``snapshot''. By doing so, those techniques could last longer and, hence, improve the performance of parallel applications.

We observed that no metric could rank load balancing techniques by their efficiency at a given iteration for a given problem. In the literature, load balancing techniques are compared using their cost, capability to correct the imbalance, and the resulting wall time of a target application on which the algorithm is applied. However, these metrics do not explain why and when a given partitioning algorithm performs well or not.  
For instance, Lieber~et~al~\cite{Lieber2018HighlyModeling} employed COSMO-SPECS+FD4 (an atmospheric simulation model) to compare their SFC-based load balancing to the Zoltan geometric partitioning methods with respect to the balancing cost, the parallel application time, and the communication cost. 
Deveci~et~al.~\cite{Deveci2016} compared their multi-jagged RCB against the classical RCB available in Zoltan with respect to their cost to balance a high number of points and their capability to produce high-quality partitions.

Finally, finding when the load balancing mechanism should be used is challenging work that has been targeted heavily over time. This challenge is usually solved using load balancing criteria that conditionally trigger the load balancing mechanism based on application data or user-defined parameters. Roughly, load balancing criteria can be divided into two categories: automatic and non-automatic. Automatic load balancing criteria, such as the one proposed by Menon~et~al.~\cite{H.MenonandN.JainandG.ZhengandL.Kale2012} or the one we proposed in a previous work~\cite{Boulmier2021OptimalCriteria}, do not require any user interventions and are backed up by theoretical models and a mathematical formulation of the criterion. In contrast, non-automatic criteria require user-defined parameters such as a desired performance improvement post-balancing~\cite{Procassini2004LoadCalculations}, an evaluation phase~\cite{Zhai2018} during which several metrics are measured, or a processing element's workload threshold~\cite{MarquezClaudio2013AApplications}. For a thorough review of load balancing criteria, we suggest the reader to refer to~\cite{Boulmier2021OptimalCriteria}.

In the present paper, we propose a first step towards what we call ``informed partitioning'' techniques. To be more specific, informed partitioning consists of using application information to drive the partitioning method to create long-lasting partitions. Herein, we illustrate informed partitioning by introducing a novel partitioning algorithm for particle simulations based on the well-known recursive coordinate bisection algorithm. In particular, we demonstrate that using the particle velocities to guide the space division improves the partition lifetime and increases the application performance. This algorithm recursively bisects the particles by their position. However, unlike the classical RCB, it bisects the domain parallel to the average velocity axis. As a result, it reduces the transfer of particles among subdomains due to their movement during the following iterations. Hence, it increases the time during which the partitioning is relevant because it reduces the load imbalance increase rate, and therefore, decreases the number of times the load balancer needs to be invoked. Ultimately, this increases the application performance. We implement a 2D version of this algorithm, and we discuss its 3D generalization while keeping its development as future work. Besides, we highlight the implementation challenges and discuss the limitations of the proposed algorithm. In a second step, we introduce a load balancing effort metric that considers both the partitioning method's ability to correct the imbalance, the cost of the load balancing operation, and the load imbalance growth. Rather than providing information about the global efficiency of the load balancing algorithm, this metric provides information about its efficiency between two consecutive load balancing calls. We show, using a numerical study on a 2D Lennard-Jones particle simulation, that our informed partitioning algorithm improves the execution of two complex examples of N-Body problems compared to the geometric partitioning techniques available in the Zoltan library~\cite{Boman2012Zoltan2}. We also demonstrate that our metric can highlight which partitioning method is the most efficient and when. Finally, we discuss the potential of informed partitioning, and we encourage researchers to try to develop new techniques that employ the idea of using application data to increase the partition lifetime. 

Section~\ref{sec:norcb} introduces our novel algorithm that implements the concept of ``informed'' partitioning.
Section~\ref{sec:perf_metric} presents our effort metric that combines all aspects of the load balancing algorithm. 
Section~\ref{sec:numerical_study} assesses the performance of informed partitioning through the comparison of our novel algorithm with respect to the geometric partitioning algorithm available in the Zoltan library. In parallel, we study the behavior of the various geometric partitioning algorithm through our effort metric. 
Section~\ref{sec:conclusion} concludes this work, discusses the potential of the concept of informed partitioning, and draws some connections with previous works. Finally, we give insight for future works.

\section{Informed Partitioning} \label{sec:norcb}

Geometric partitioning techniques are algorithms that divide the computational domain into sub-domains using geometric shapes and by considering the spatial coordinates of the computational elements. In contrast, Graph partitioning techniques focus on dividing the graph formed by the computational elements (vertices) and their interactions (edges) while minimizing the edge cuts. Herein, we introduce the concept of informed partitioning via a geometric partitioning proof-of-concept for N-Body simulations.

In classical geometric partitioning techniques, the algorithm typically considers a snapshot of the positions of the computational elements. Then, using geometric constructions, the algorithm divides the space into subdomains, focusing on providing a high load imbalance correction capability at a low CPU time cost. For instance, some researchers proposed to use Voronoï tessellations~\cite{Zhakhovskii2005ASimulation,Fattebert2012,Egorova2019ParallelSubdomains}, some divided the space into cuboids~\cite{Begau2015}, or improved classical algorithms such as RCB~\cite{Simon1997HowBisection} by allowing multisections instead of bisections~\cite{Deveci2016}. However, it remains unclear why a given way of dividing the space is better than others, provided that they have the same cost and load imbalance correction capability. Note that graph partitioning techniques like METIS~\cite{Karypis1998AGraphs} are also popular load balancing techniques but are less used in particles simulations due to their complexity. In this paper, we focus on geometric partitioning techniques.

We remarked in a previous paper that besides their cost and their capability to correct the imbalance, partitioning methods change the load imbalance growth because they modify the shape/size of subdomains~\cite{Boulmier2021OptimalCriteria}. Hence, to improve partitioning techniques, one must focus on reducing these three aspects altogether: (i) cost of load balancing, (ii) capability to correct imbalance, and (iii) load imbalance growth. In particular, we remarked that the last aspect is not much considered in the literature even though we will see in the numerical study that this aspect can significantly impact the performance. 

To reduce the load imbalance growth, one needs to reduce the number of computational elements that migrate between processing elements at each iteration. Note that this is a problem only if the computational elements ultimately gather into a subset of the processing elements, thus creating load imbalance. In particles simulations, the reason for this migration lies in the particle movement according to their velocity. Indeed, each subdomain possesses an $(+)$~ingoing and $(-)$~outgoing particles flow (depending on how many particles move in or out of their own space), if the sum of these flows is positive then the processing element is currently loading. The problem of load imbalance arises when some subdomains have a positive particle flow while being balanced (i.e., the particle converges to a few subdomains). Therefore, by reducing the number of migrating computational elements, we reduce the outgoing flows and also the load imbalance increase rate. Note that not every particles simulation suffers from load imbalance created by the particle migration. For example, as suggested by~\cite{Hirschmann2016}, in Brownian motions, the migration of particles does not create load imbalance over time because there are no preferred direction for the particles. Hence, there is no particular gathering of particles in a subset of processing elements. 

To reduce particles migration, one could use the velocity of the bodies to predict where they will move and change the partitioning accordingly (e.g., by lengthening the subdomain in the direction of the trajectory of the bodies). Likewise, one could analyze the forces applied to the bodies to drive the creation of the partitions. \begin{figure}
    \centering
    \includegraphics[width=0.5\textwidth]{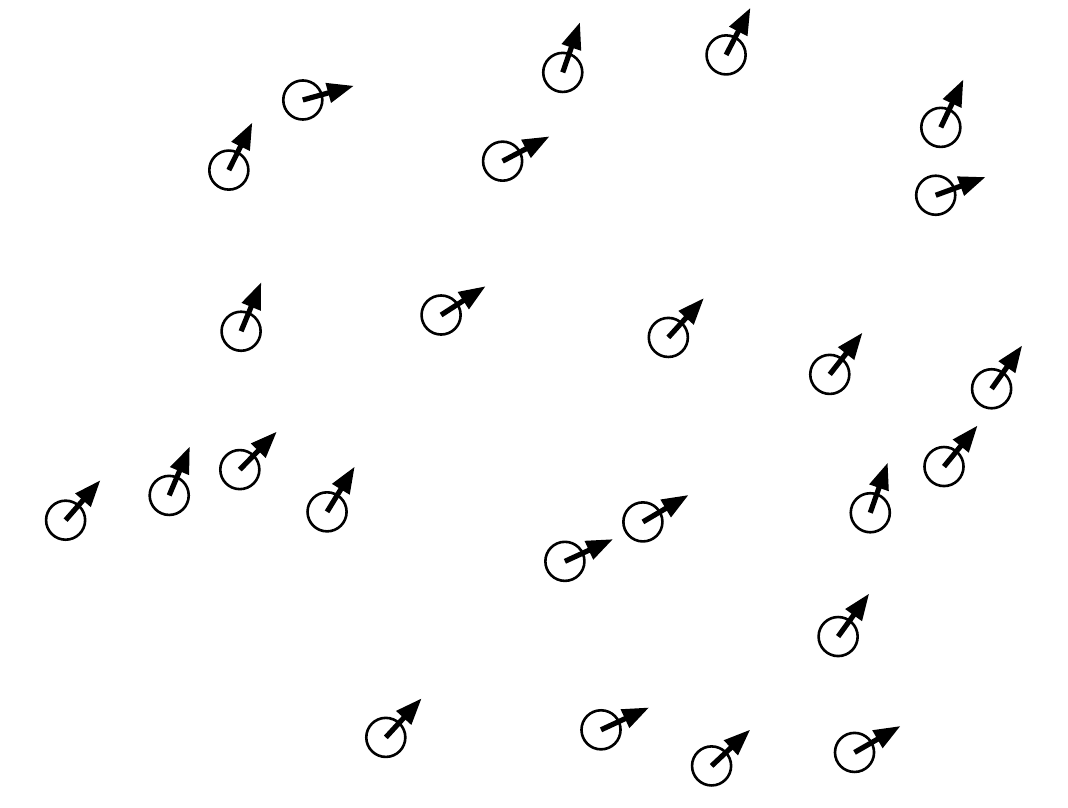}
    \caption{$25$ particles randomly arranged in a square. The arrow indicates the direction in which each body moves.}
    \label{fig:toy}
\end{figure} We illustrate this using a simple example in Figure~\ref{fig:toy}. Therein, $25$~particles are randomly placed in a square. Now, let imagine that we want to partition this domain into four load balanced subdomains. A straightforward way would be to divide the space into four vertical stripes containing a quarter of the particles. However, it is clear that this does not take into account how the bodies move, and eventually, this partitioning will require frequent load balancing calls. Instead, using information from physics, we could roughly predict that the particles are going into the upper right corner. Hence, we could partition according to this information to lower the migration of elements. We call the process of using application data to drive the domain partitioning to increase the partition lifetime: ``informed partitioning''. In Figure~\ref{fig:intro_norcb_vs_rcb}, we give a concrete example of a partitioning that takes into account the movement of the bodies and one that does not.

\subsection{Non-Orthogonal Recursive Bisection}

Herein, we propose to implement the idea of informed partitioning using a bisection of the computational space in the axis formed by the average velocity at the median workload point, producing non-orthogonal bisections. We call this algorithm the Non-orthogonal Recursive Coordinate Bisection (NoRCB). In contrast, the traditional recursive coordinate bisection applies orthogonal cuts at the median workload points. NoRCB repeats the bisection process recursively until $P$ subdomains have been produced. By using the particle velocities to guide the bisection axis, we reduce the load imbalance that arises from the migration of the particles across processing elements. \begin{figure}
    \centering
    \includegraphics[width=0.8\textwidth]{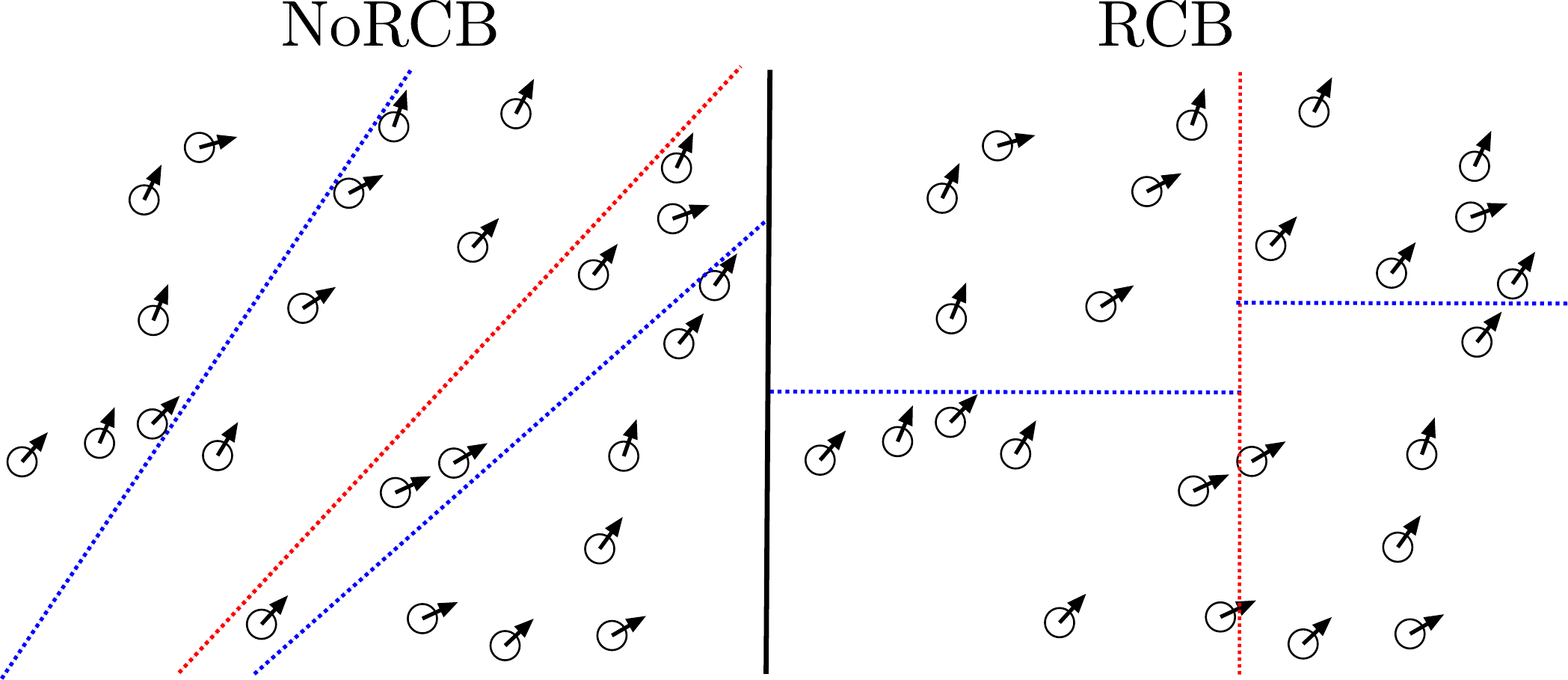}
    \caption{Toy example with $25$ particles and their associated velocity (arrow) comparing the non-orthogonal recursive coordinate bisection (NoRCB) and the classical recursive coordinate bisection (RCB) with respect to their resulting partitioning. The red dashed line corresponds to the first bisection, whereas the blue dashed lines correspond to the second and third bisection. In the right figure, RCB bisects the domain along the longest axis at the median point. NoRCB bisects the computational domain recursively along the axis formed by the average velocity vector at the median point in the left figure. The load imbalance growth is expected to be smaller on the left figure due to less particle migration over time. Note that NoRCB will naturally produce a stripe partitioning the particles go approximately in the same direction}
    \label{fig:intro_norcb_vs_rcb}
\end{figure} In Figure~\ref{fig:intro_norcb_vs_rcb}, we illustrate the differences and the motivation for our algorithm and, in particular, the benefits of informed partitioning. This figure shows two identical sets of particles with similar velocities on which informed partitioning (NoRCB) and RCB are applied. On the left figure, we observe that because the bisections follow the average velocity, the particles within the domain are less prone to cross the processing element boundaries. Therefore, in addition to correcting the load imbalance, we reduce its growth over the forthcoming iteration. This contrasts with the traditional RCB partitioning, in which we observe that many particles will migrate between subdomains after few iterations. \begin{figure}
    \centering
    \includegraphics[width=0.8\textwidth]{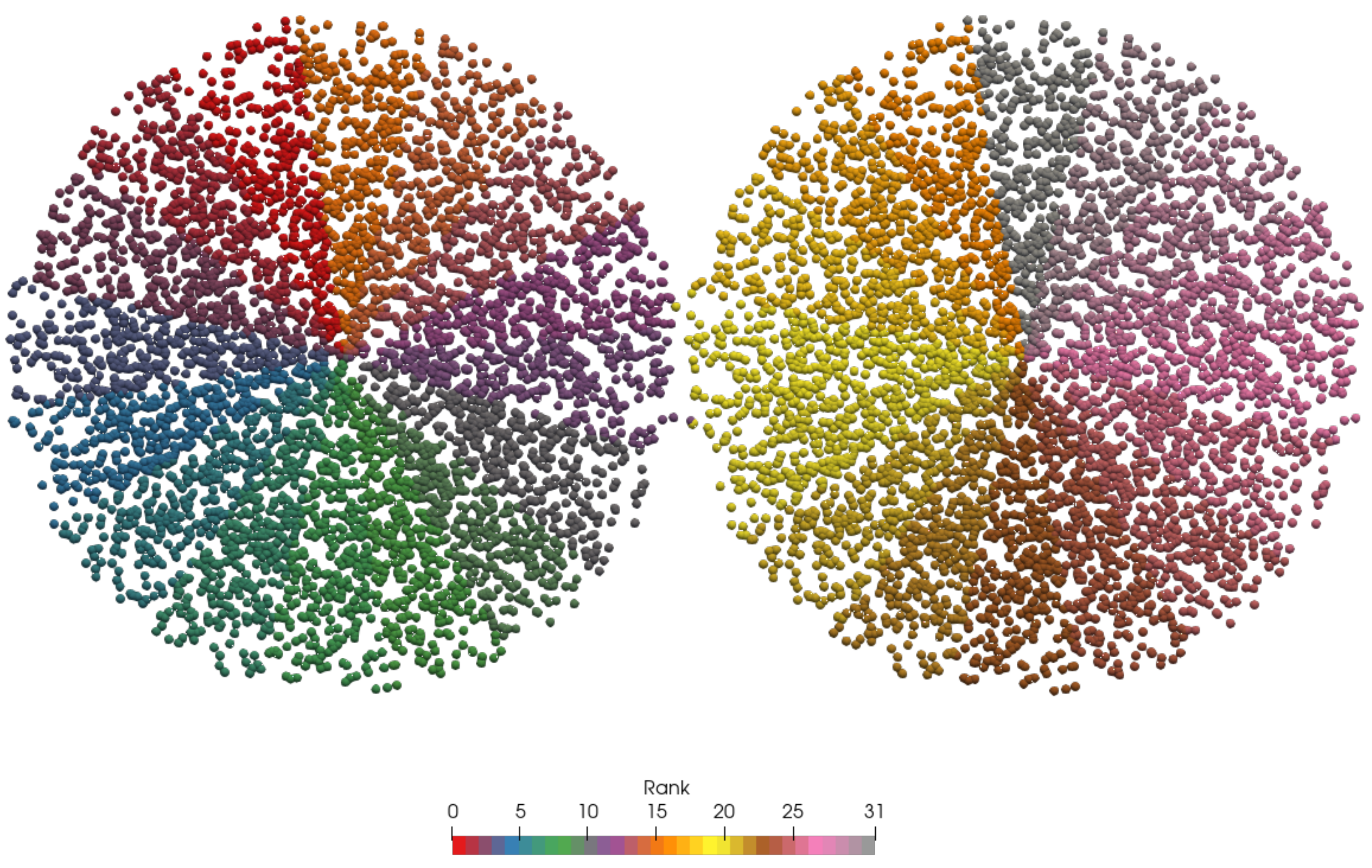}
    \caption{Partitioning of two gas disks attracted toward their center using ``informed partitioning'' (NoRCB) on 32 processing elements. The particle color corresponds to the rank of their attributed processing element.}
    \label{fig:contracting_gas_disks}
\end{figure} To further demonstrate the partitioning produced by our algorithm, we show in Figure~\ref{fig:contracting_gas_disks} the result of NoRCB partitioning with $32$ processing elements on two gas disks that contracts. 

The first step in our novel algorithm consists of computing the average velocity vector $\mu_v$ of the particles belonging to the current domain $D$. This vector will be used as the cutting axis. Then, we obtain the angle $\alpha$ between $\mu_v$ and the Y-axis. Next, we rotate the computational elements by $\alpha$ degrees such that the average velocity of the rotated elements is pointing upward. It allows us to find the median particle on the X-axis using a classical parallel selection algorithm, such as the Quickselect~\cite{Hoare1961AlgorithmFind}. Afterward, the median element $m$ is employed to produce two sub-domains $s^{<=}$ and $s^>$ containing the elements below or equal and greater than $m$, respectively. Finally, we apply the inverse rotation (i.e., $-\alpha$). At this point, the initial domain $D$ is divided into two equally sized subdomains. The bisection can continue recursively with $D=s^{<=}$ and $D=s^{>}$, until $P$ equally sized subdomains have been created. The algorithm pseudocode is presented in Algorithm~\ref{alg:norcb-pseudocode}.
\begin{algorithm}\small
\SetKwProg{Fn}{Function}{}{end}\SetKwFunction{FRecurs}{NonOrthogonalRecursiveBisection}%
\newcommand{\forcond}{$i=0$ \KwTo $n$}
\KwData{$D$: the domain to bisect \\ N: the number of elements in $D$ (i.e., $\mid D \mid$)\\ P: the number of processing elements}
\KwResult{$S^*$: The subdomains such that $\bigcup\limits_{s \in S^*} s = D$ and $\mid s \mid \approx \frac{N}{P} \ \forall s\in S^*$}
$S^* \leftarrow \{D\}$\;
\While{$\mid S^* \mid < P$} {
    $S = \emptyset$\;
    \ForEach{subdomain $s \in S^*$ } {
        $\vec{\mu_v} \leftarrow$ average velocity in $s$\;
        $\alpha \leftarrow$ angle between $\vec{\mu_v}$ and the Y-axis\;
        rotate the elements in $s$ by $\alpha$ degrees\;
        $m \leftarrow$ median element in $s$\;
        split $s$ into $s^{<=}$ and $s^>$ containing elements respectively below or equal and greater than $m$\;
        split the 2D or 3D domain into two sub-domains\;
        rotate back the elements in $s^{<=}$\;
        rotate back the elements in $s^{>}$\;
        push back $s^{<=}$ and $s^>$ into $S$\;
    }
    $S^* = S$\;
}
\Return{$S^*$}
\caption{Pseudo-code of the Non-Orthogonal Coordinate Bisection algorithm.}
\label{alg:norcb-pseudocode}
\end{algorithm}

\paragraph{2D implementation} In $2$~dimensions, the domains are planar and thus are bisected by a line. Hence, only the average velocity vector and the median point are required to apply our bisection algorithm. Therefore, particular attention must be given to the average velocity computation and the median finding algorithm. 

First, one may want to avoid using a null average velocity vector in the NoRCB algorithm as it will raise errors due to division by zero. A null average velocity vector can be found in Brownian motion simulations, and it indicates that there is no preferred directions for the particles. Therefore, the average velocity computation must handle such a situation properly. 

The average velocity computation can be done in multiple ways. The most straightforward way is, of course, to average the velocity among all particles. However, one could also sample the particles and compute the average velocity among the samples to speed up the computation. Another trick could be to consider each velocity vector's supporting line without considering each particle's ``direction'' (i.e., velocity sign). Each of these methods seems to provide pros and cons and could improve the partitioning lifetime. However, this is out of the scope of this paper and will be targeted in future works. Herein, we used the most straightforward method that consists of averaging the velocity of all particles, and we discuss its limitations in Section~\ref{subsec:limitations}.

The median finding algorithm is a crucial part. It defines the balancing capability of our algorithm. In other words, the load balancing capability is directly related to the ability of the median finding algorithm to determine the median point. For instance, if it yields a point that splits the points into $30\%$ below and $70\%$ above, not only the current partitioning is terrible, but this imbalance will be propagated due to the recursive nature of the coordinate bisection. In our algorithm, we used a parallel version of the quickselect~\cite{Hoare1961AlgorithmFind} algorithm to find the median point. We differ from the Zoltan~\cite{Devine2002} RCB implementation that uses a binary search approach. Therein, they divide the space recursively using multiple cuts until they find an approximation of the median that satisfies a tolerance criterion.

In contrast, our approach finds a pivot among the data. Then, it removes the data above or below the pivot if the median does not lie in the subset. Finally, the operation is repeated until the pivot is the median. Nevertheless, we observe that our parallel implementation needs to be optimized, which we will in future work. This future paper will focus on proposing an optimized implementation of this novel algorithm and integrating it within the Zoltan2~\cite{Boman2012Zoltan2} framework. At this stage, the current paper focuses on introducing the idea of informed partitioning (i.e., using the physics data to drive the partitioning algorithm) and showing a proof-of-concept of this new idea. 

\paragraph{3D implementation} In $3$~dimensions, the domains are bisected by plans, which has one more degree of freedom than a line. The general procedure remains the same: finding the average velocity vector and computing the median point. However, the third dimension provides an extra freedom point to define the bisecting plan properly. We call this point the rotation point because it defines the rotation of the plan around the line supported by the velocity vector and the median point. It is unclear if every point is equivalent with respect to the partition lifetime. \textit{A priori}, the partition quality is not affected by the choice of the rotation point because it does not alter the quality of the median finding algorithm. However, the rotation point may be necessary, lifetime and communication-wise. For instance, setting the rotation point in the middle of a particle cluster may create much communication between processing elements, or it may increase the probability that particles migrate from a processing element to another. It is why further research efforts are required before implementing this algorithm in 3D. Hence, we let this task for future work.  

\subsection{Limitations}\label{subsec:limitations}
The first caveat of the proposed algorithm is that the number of neighbors of each subdomain is unbounded like the classical RCB algorithm. Also, because the algorithm cuts the space non-orthogonally, it may create slender subdomains with more neighboring processing elements. Unfortunately, this limitation is inherent to the tessellation produced by the algorithm and can not be improved without restricting the bisection freedom (e.g., rounding the average velocity).

The second limitation is related to the average velocity vector computation. Indeed, one may want to avoid summing opposing velocities as it may result in a null velocity vector (in the case of uniformly distributed velocities with zero mean, for instance). In such cases, when averaging these velocities, the resulting vector may be null or very small. It indicates that there is no preferred cutting direction and, as such, the bisection should be orthogonal to the longest axis (X or Y) to minimize the surface of communication between processing elements. In our algorithm, we verify that the norm of the average velocity vector is at least greater than a given small threshold; otherwise, we set the average velocity vector such that it is orthogonal to the longest axis. Currently, the threshold value must be user-defined and, herein, we used the value $1e-3$, which we found to work well in our experiments. Moreover, one of our hypotheses is that this threshold should depend on the cut-off radius. Indeed, if we know that the average velocity is less than the speed required by a particle to travel X percent of the cut-off radius (for instance), then there is only a small amount of particles that may migrate in the forthcoming iterations. Nevertheless, the study of this threshold and how to find a proper value for it is let for future work. 

Finally, our implementation of QuickSelect can be optimized, especially in the case of data already partitioned but imbalanced into processing elements. However, we plan to implement a more scalable and efficient parallel selection algorithm such as the one proposed by Siebert~\cite{SiebertScalableAndEfficient2014} as well as optimizing the parallel implementation of our algorithm. Note that this does not affect the message we are trying to give throughout this paper: informed partitioning is a way to increase the partition lifetime to improve application performance. Moreover, as we will see in the numerical study, our non-optimized implementation is capable of being more efficient than the RCB, RIB, and HSFC implementation of the Zoltan~\cite{Devine2002} library on two experiments with up to $128$ processing elements. 

\section{An Effort Metric for Load Balancing Techniques} \label{sec:perf_metric}

In this section, we are interested in ranking load balancing algorithms by the effort they cause on the application at an arbitrary iteration without executing the whole application. The effort is measured by the amount of work done between two load balancing calls. It is essential to understand that this is different from looking for the algorithm that corrects the most imbalance or is the cheapest to correct the imbalance. Indeed, two algorithms may have the same cost, but one may produce way more load imbalance over time or need more frequent load balancing. Herein, we want to combine all various aspects of load balancing techniques that affect the application performance into one metric.

In a previous work~\cite{Boulmier2021OptimalCriteria}, we studied the time for a load-balanced parallel application with a recurring load imbalance pattern. Therein, we proposed an equation inspired by the work of Menon~et~al.~\cite{H.MenonandN.JainandG.ZhengandL.Kale2012} that computes the parallel application time after $\gamma$ iterations and load balanced every $\tau$ iterations. This equation reads
\begin{equation} \label{eq:tpar_1}
    T_{\text{par}} = \sum_{i=0}^{n-1} \big( \int_{0}^{\tau} u(x) \text{d}x + C \big) + \int_0^\gamma \mu(x) \ \text{d}x.
\end{equation} where $n$ is the number of times the load balancing algorithm is used, $u(\cdot)$ is the load imbalance given by the imbalance time metric~\cite{DeRose2007DetectingSystems}, $C$ is the load balancing cost, and $\mu(t)$ is the average workload. The imbalance time metric measures the difference in time between the slowest processing element and the average iteration time, and it reads
\begin{equation}\label{eq:derose}
    u(t) = \text{max}(t) - \mu(t), 
\end{equation}, where $t$ is an iteration and max$(t)$ is the time of the slowest processing element at iteration $t$. Also, we remarked that this equation is strongly dependent on the chosen load balancing algorithm, which affects $C$, $\tau$ (provided that an automatic and theoretically optimal load balancing criteria is used), $u$, and therefore $T_{\text{par}}$ itself. To recall, $u$ and $\tau$ depends on how well the partitioning technique fits with the problem to solve, while $C$ is directly related to the complexity of the partitioning method and the migration time of the computational elements.  
Hence, let $\mathcal{A}$ be the set of the $n$ available load balancing algorithms $a_1, a_2, ..., a_n$, and $T_{\text{par}}^{a}$ be the parallel time of the current application, load balanced using algorithm $a$. Surely, the optimal algorithm $a^*$ is
\begin{equation}
    a^* = \argmin\limits_{a \in \mathcal{A}} T_{\text{par}}^{a}.
\end{equation} However, this is of no use during application execution as it requires the application to be fully executed. Furthermore, it does not give any information on when the particular algorithm is efficient or not. For that purpose, we propose to measure the average time-per-iteration during load balancing intervals. Note that this does not guarantee that the algorithm is globally optimal, but it gives local information about the algorithm efficiency. Moreover, assuming a principle of persistence~\cite{LaxmikantV.Kale2002}, a locally optimal technique may remain optimal during a few load balancing intervals, which could justify the use of such a metric for load balancing algorithm selection. In the next section, we will confirm this hypothesis with the numerical experiments. 

Now, let us decompose Equation~\ref{eq:tpar_1} into $t$ balancing intervals respectively at $\tau_0, \tau_1,...,\tau_{t-1}$. Assuming that a load balancing is done at iteration $0$, then $\tau_0 = 0$. Besides, we underline that no load balancing is done at the end of the application execution. This reads
\begin{multline}
\label{eq:tpar_a}
    T_{\text{par}}^{a} =  \int_{0}^{\tau_1^a} u_0^a(x) \text{d}x + C_0^a +
                      \int_{\tau_1^a}^{\tau_2^a} u_1^a(x) \text{d}x + C_1^a + ... +
                      \int_{\tau_{t-1}^a}^{\gamma} u_{t-1}^a(x) \text{d}x + C_{t-1}^a +\\ \int_0^\gamma \mu(x) \text{d}x.
\end{multline} 
Surely, it is easy to compare two algorithms having the same intervals. The current best algorithm is the one that minimizes the time of the current balancing interval. However, in practice, load balancing intervals have different sizes and thus are more complicated to compare. Hence, to evaluate the efficiency of a load balancing algorithm $a$ during the interval $[\tau_{i}^a, \tau_{i+1}^a]$, we propose to compute the average balancing interval time-per-iteration $\hat{T}_{\tau_{i}}^a$, which reads 
\begin{equation}
\label{eq:ttau_a}
    \hat{T}_{\tau_i}^a = \frac{\int_{\tau_{i}^a}^{\tau_{i+1}^a} u_{i}^a(x) \text{d}x + C_{i}^a}{\tau_{i+1}^a - \tau_{i}^a}.
\end{equation} Using this equation, it is trivial to compare the efficiency of two algorithms at any point during application execution. For instance, let $k$ be the iteration at which we want to compare algorithm $a$ and $b$. Also, \mbox{$\tau_{i}^a \leq j \leq  \tau_{i+1}^a$} and \mbox{$\tau_{j}^b \leq k \leq  \tau_{j+1}^b$.} Hence, $a$ performs better than $b$, during their load balancing interval that includes iteration $k$, if
\begin{equation*}
    \hat{T}_{\tau_i}^a < \hat{T}_{\tau_j}^b.
\end{equation*} 

Nevertheless, even if comparing two algorithms becomes easy using the effort metric. One limitation lies in the lack of information about what aspect of load balancing generates most of the effort. Indeed, it is still unclear if a given effort comes from the load balancing cost $C$, the imbalance correcting capability, or the speed of the load imbalance growth. Having a metric with a high level of explainability could be of great use to better understand why some partitioning methods perform better than others. A possible solution could be to estimate the ratio of each component of the load balancing effort. However, this is out of the scope of this paper and will be targeted for future work.

\section{Numerical Study} \label{sec:numerical_study}
In this section, we assess the performance of the proposed informed partitioning concept. First, we present a simple toy example showing that NoRCB effectively reduces the load imbalance growth, which significantly helps in reducing the application wall time. Afterward, we compare the parallel time of our implementation of informed partitioning against the geometric partitioning techniques (RCB, RIB, and HSFC) available in the Zoltan~\cite{Devine2002} framework on three 2D N-body simulations exhibiting various particle configurations. The partitioning has been applied to the positions of the particles. The experiments have been executed with YALBB~\cite{XetqL/yalbb:Benchmark}, a homemade load balancing benchmark based on N-body simulations with a short-range force. In particular, the Lennard-Jones potential is used. 

In the toy example, we condense a gas disk comprising $10{,}000$ particles during $5{,}000$ iterations. Note that this experiment is particularly suited for our implementation of informed partitioning because the movement of the particles is not curved (at least during some iterations). Hence, the bisection follows their movement perfectly, and it reduces the load imbalance growth drastically. Therein, to simplify the load imbalance analysis, we consider a periodic load balancing criterion that triggers the load balancing mechanism every $600$ iterations. The goal is to assert that the imbalance cumulated over the load balancing period grows slower with NoRCB than with the other techniques. However, doing so with an automatic load balancing criterion is much more difficult because such a criterion minimizes the global load balancing effort. Therefore, it is hard to observe if the load imbalance growth is effectively reduced compared to the other aspects.
\begin{figure}
    \begin{center}
    \subfloat[Parallel time\label{fig:tpar_toy}] {
    \includegraphics[width=.49\linewidth]{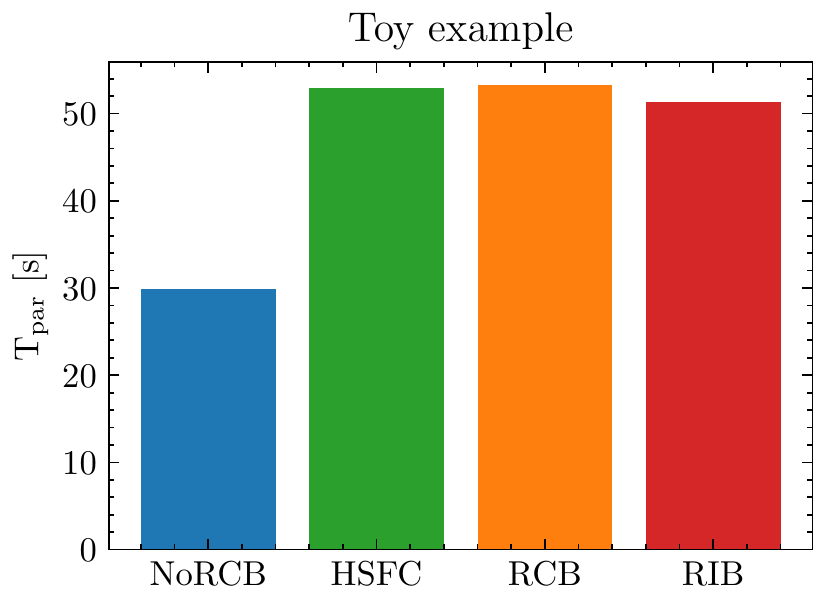}
    }
    \subfloat[Cumulated imbalance\label{fig:cum_imb_toy}]      {
    \includegraphics[width=.49\linewidth]{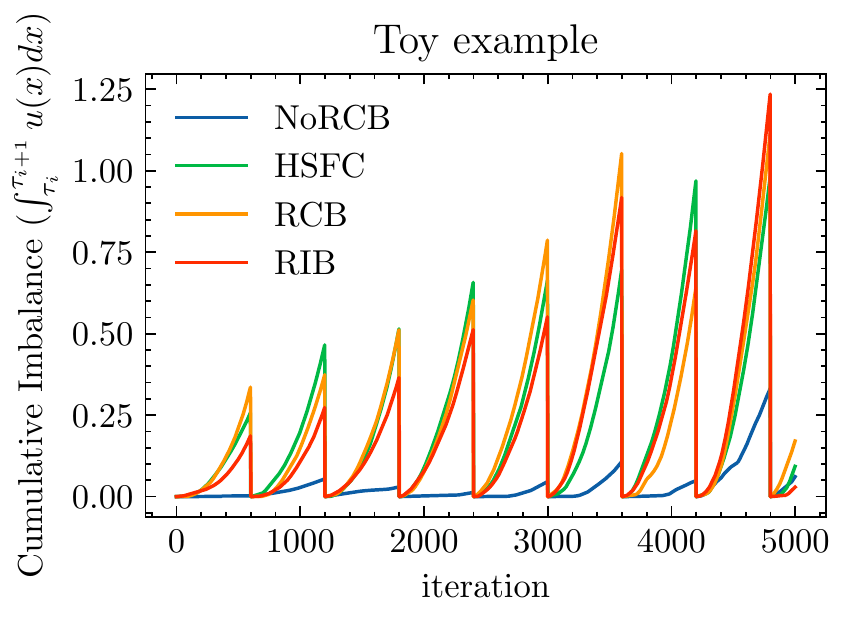}
    }
    \end{center}
    \caption{Toy example that contracts a gas disk of $10{,}000$ particles with a load balancing step every $600$ iterations. The right figure shows the cumulative load imbalance over time. The left figure indicates the parallel time of each load balancing algorithm. NoRCB effectively reduces the rate at which the load imbalance is created compared to classical geometric partitioning techniques, which greatly improves the parallel application time.}
    \label{fig:numerical_toy}
\end{figure} 
In Figure~\ref{fig:tpar_toy}, we observe that NoRCB is almost twice as fast as the other techniques. Indeed, the reason is given in Figure~\ref{fig:cum_imb_toy}, where we observe that within each load balancing interval of $600$ iterations, the load imbalance grows significantly slower with NoRCB. It confirms that informed partitioning can reduce the load imbalance growth. 

Now, we go further in the assessment of our concept by testing it in the three complex experiments. Therein, we employed the theoretically optimal and automatic load balancing criterion that we introduced in~\cite{Boulmier2021OptimalCriteria}. This criterion triggers the load balancing mechanism when \begin{equation*}
    \tau u(\tau) - \int_0^\tau u(x)\text{d}x = C,
\end{equation*} is satisfied. To recall, $\tau$ indicates the load balancing interval, $u(\cdot)$ is the load imbalance time metric from DeRose~et~al.~\cite{DeRose2007DetectingSystems} (see Equation~\ref{eq:derose}), and $C$ denotes the load balancing cost. This balancing criterion is derived from Equation~\ref{eq:tpar_1} and perform better than the criterion form Menon~et~al.~\cite{H.MenonandN.JainandG.ZhengandL.Kale2012} in most of our test cases. 

\begin{figure}
\centering
\begin{minipage}{.48\textwidth}
  \centering
  \includegraphics[width=1.0\textwidth]{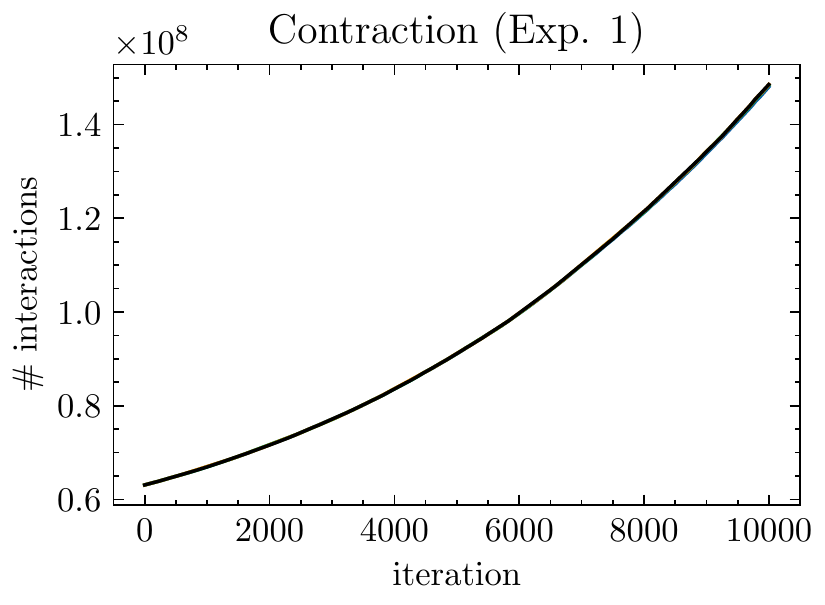}
  \captionof{figure}{A gas disk of $40{,}000$ uniformly distributed particles is contracting. The density increases as the particles move towards the center of the disk and the number of interactions to be computed increases as shown in the plot.}
  \label{fig:inter_disk}
\end{minipage}\quad
\begin{minipage}{.48\textwidth}
  \centering
  \includegraphics[width=1.0\textwidth]{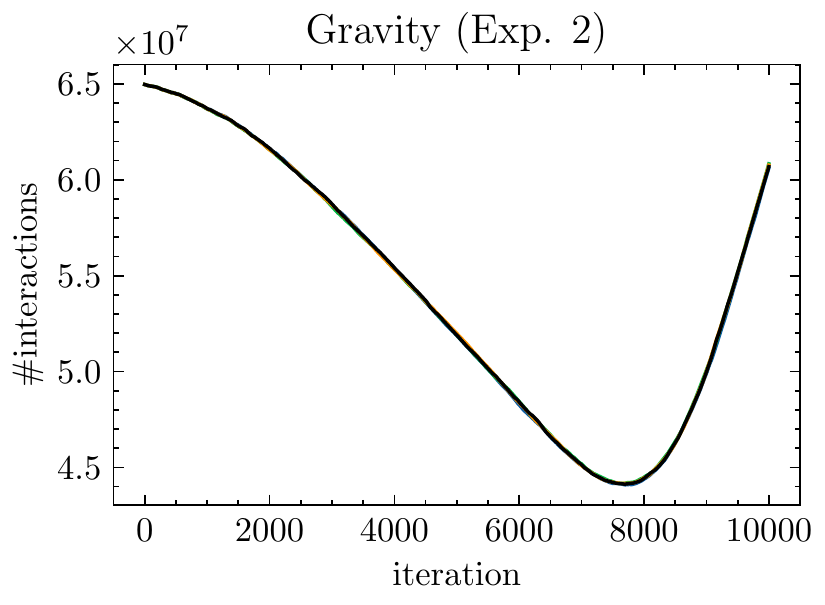}  
  \captionof{figure}{A half square gas containing $40{,}000$ uniformly distributed particles is subject to the gravity. The density decreases as the particles are attracted towards the ground and increases again when they reach it, hence a variation of the workload resulting from the change of interacting pairs of particles.}
  \label{fig:inter_gravity}
\end{minipage}
\begin{minipage}{.48\textwidth}
  \centering
  \includegraphics[width=1.0\textwidth]{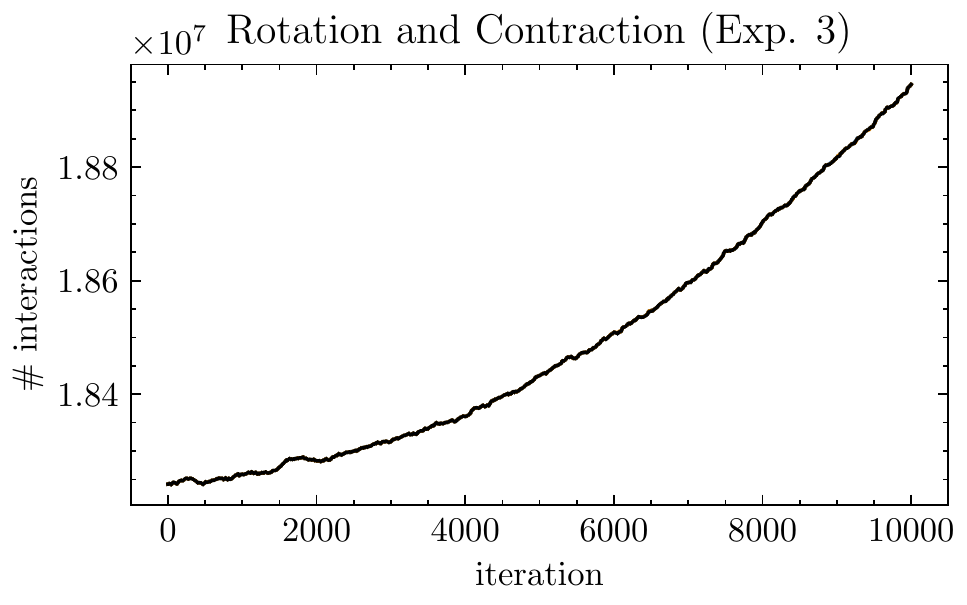}  
  \captionof{figure}{A rotating gas disk of $10{,}000$ uniformly distributed particles is subject to an attraction force towards its center. The density increases as the disk contracts under the attraction force. As in the Contraction experiment, the number of interaction to compute grows as the gas disk contracts.}
  \label{fig:inter_blackhole}
\end{minipage}
\end{figure}

The first experiment, ``Contraction'', reproduces the toy example with more particles. It condenses a gas disk composed of $40{,}000$ particles. Similarly, the gas disk undergoes an attraction force towards its center. To illustrate the complexity of this task, Figure~\ref{fig:inter_disk} shows the number of interactions that must be computed at each iteration.   

The second experiment, ``Gravity'', deals with a gas composed of $40{,}000$ uniformly distributed in a rectangle of size $(0.5, 1.0)$ inside a square of size $(1.0, 1.0)$. The particles have their velocity sampled from a uniform distribution. Moreover, the gas is under a force that attracts the particles to the bottom of the square domain. This experiment is more difficult for the non-orthogonal recursive coordinate bisection (NoRCB) because the particle movement starts from a Brownian motion to a linear movement directed towards the bottom of the domain. 
Figure~\ref{fig:inter_gravity} illustrates the number of interactions to be computed at each iteration.

The third experiment, ``Rotation and Contraction'', involves a uniformly distributed gas disk with a center $c$ where each particle has a velocity perpendicular to the vector directed from the particle to $c$.  Moreover, the gas disk is attracted to the center $c$. This experiment is the hardest for informed partitioning because of the presence of the disk's angular velocity. 
Figure~\ref{fig:inter_blackhole} shows the number of interactions to compute at each iteration.

\begin{table}[ht]
\begin{adjustbox}{center}
\begin{tabular}{lccc}
\toprule
 Parameter                & Gravity & Contraction & Rotation and Contraction \\
\midrule
Domain size (x, y) & \multicolumn{3}{c}{($1.0$, $1.0$)}\\
Number of particles & $40{,}000$& $40{,}000$& $10{,}000$\\
\bottomrule
\end{tabular}
\end{adjustbox}
\caption{Domain size and number of particles for the three numerical experiments.}
\label{tab:numerical:phyparams}
\end{table}

These experiments have been conducted with $128$ processing elements (Intel~E5-2630V4, $2.2$~Ghz) on Yggdrasil, the cluster of the University of Geneva, except for the toy example that has been executed with $64$ processing elements. Each of them has been done $5$~times with each load balancing technique, and we report their median parallel time. Moreover, we show, for each load balancing algorithm, the value of our effort metric $\hat{T}_{\tau_i}$ over time and identify at what time during the application our novel technique is efficient/inefficient. The domain size and the number of particles used in the experiments are shown in Table~\ref{tab:numerical:phyparams}.

\begin{figure}[H]
\begin{center}
\subfloat[Contraction\label{fig:Tpar_Contraction_exp_1}] {
\includegraphics[width=.33\linewidth]{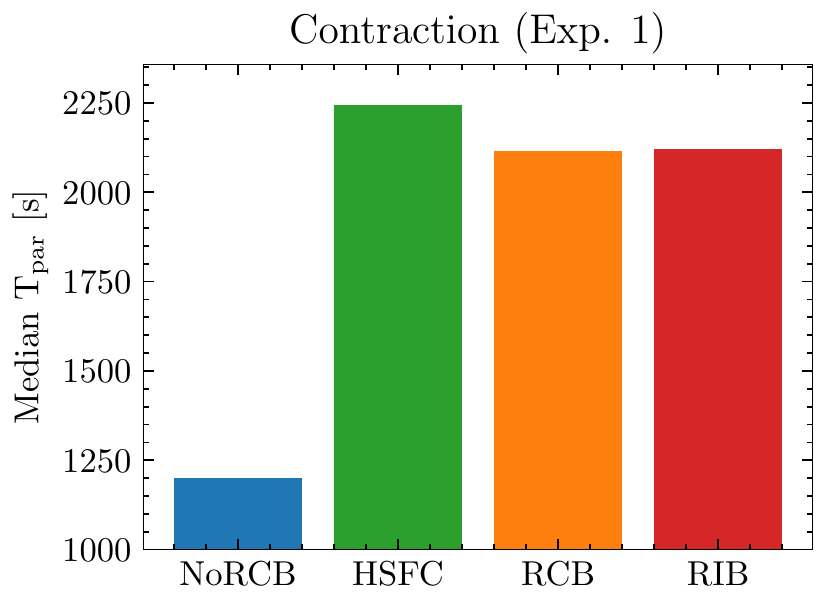}
}
\subfloat[Gravity\label{fig:Tpar_Gravity_exp_2}]      {
\includegraphics[width=.33\linewidth]{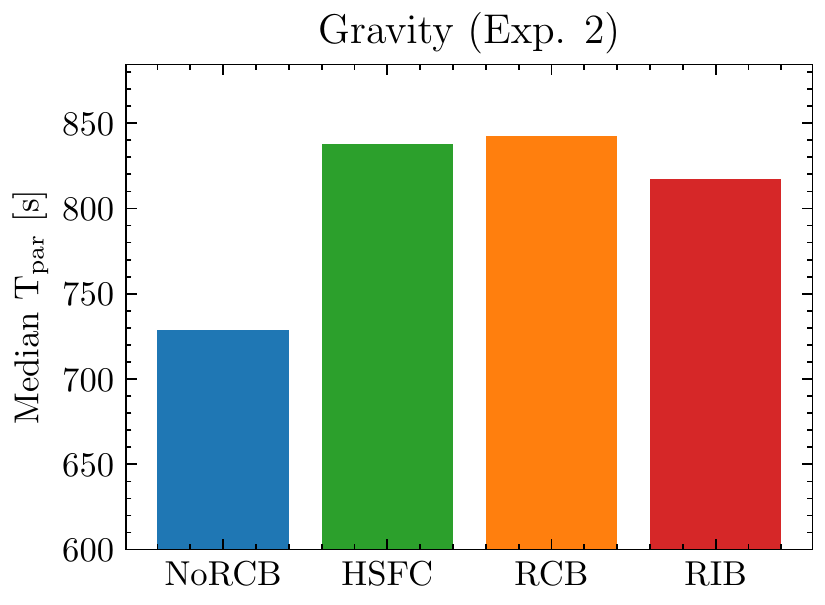}
}
\subfloat[Rotation and contraction\label{fig:Tpar_Rot_and_contract_exp_3}]      {
\includegraphics[width=.33\linewidth]{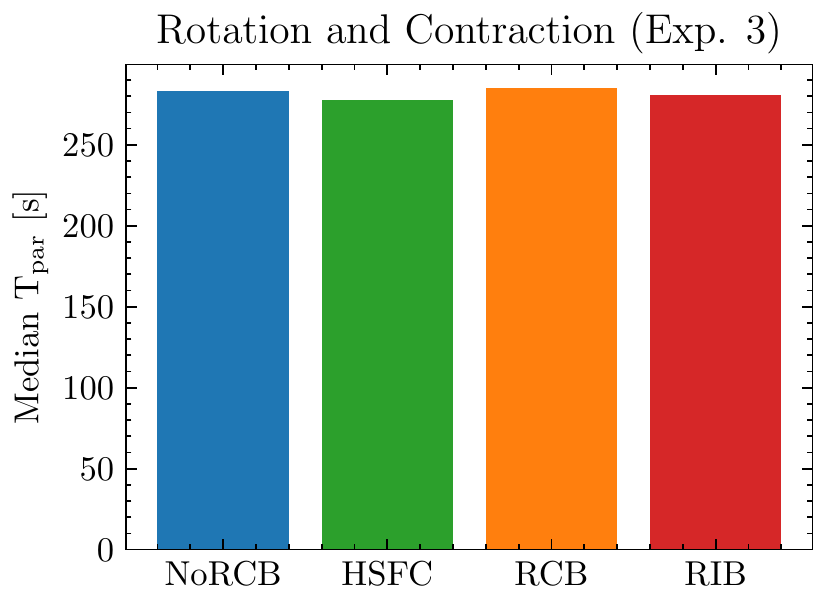}
}
\end{center}
\caption{Median parallel time, among $5$ trials, of all experiments executed with each load balancing algorithm. Our implementation of ``informed partitioning'' works best when the movement of the particles follows a straight line (Contraction experiment and the second part of the Gravity experiment). Consequently, it is less efficient in the beginning of the Gravity experiment before the particles are moving towards the ground.  ``Informed partitioning'' performs almost as good as Zoltan's geometric partitioning technique when the particle movements are curved like in the Rotation and Contraction experiment.}
\label{fig:median_tpar}
\end{figure}
\begin{figure}[H]
\begin{center}
\subfloat[Contraction\label{fig:metric:contract}] {
\includegraphics[width=.33\linewidth]{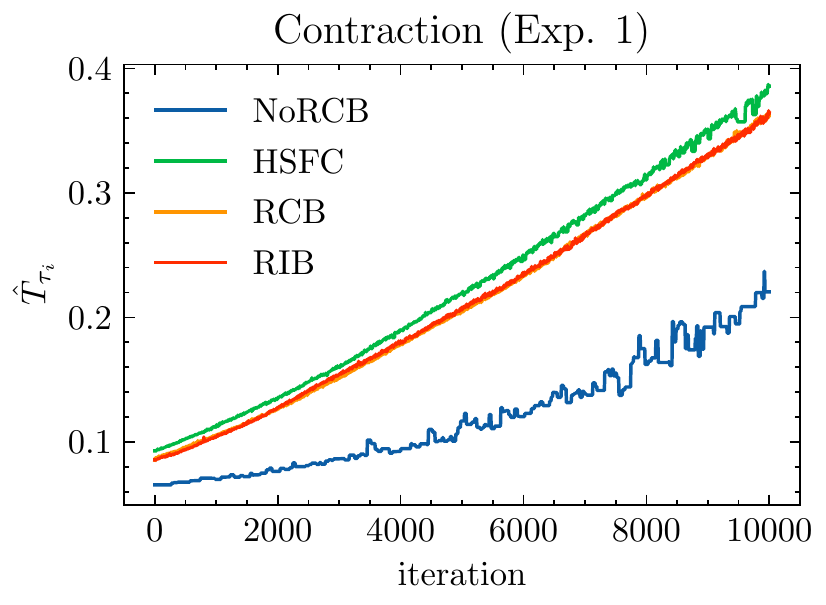}
}
\subfloat[Gravity\label{fig:metric:grav}]      {
\includegraphics[width=.33\linewidth]{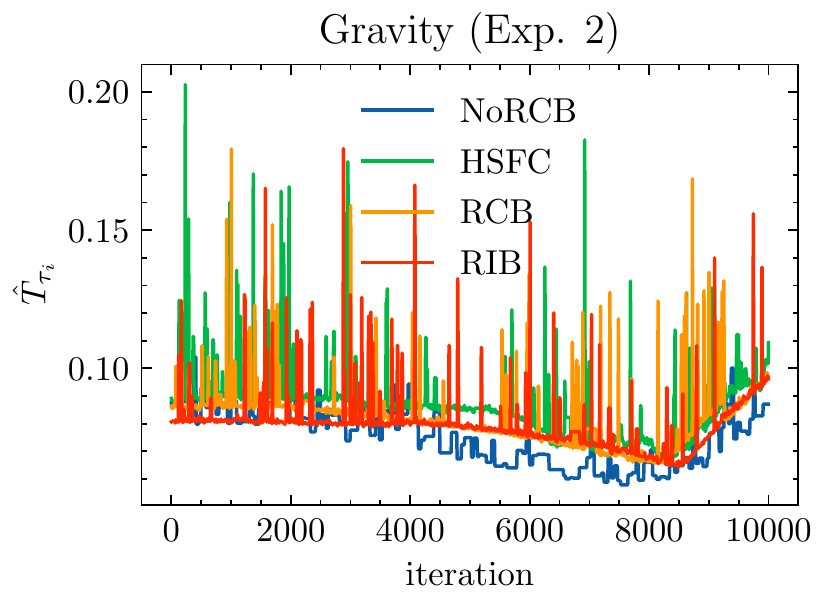}
}
\subfloat[Rotation and contraction\label{fig:metric:rotcontr}]      {
\includegraphics[width=.33\linewidth]{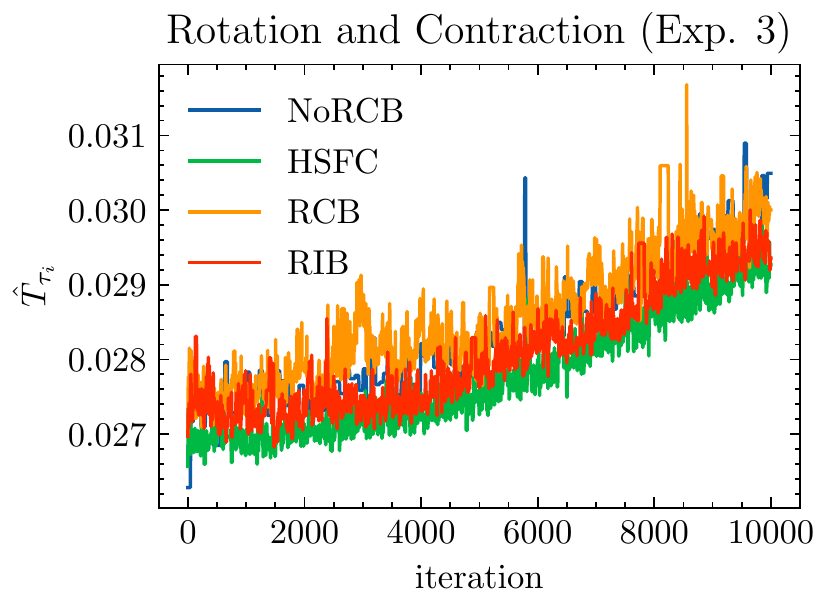}
}
\end{center}
\caption{Plot of Equation~\ref{eq:ttau_a} (effort metric) over time for each experiment. In (1), we observe that our implementation of ``informed partitioning'' (NoRCB) is the most efficient throughout application execution. In (2), we see that NoRCB starts to become more efficient when the particles have a straight movement starting from iteration $3{,}000$. In (3), all the techniques lead to an almost similar effort.}
\label{fig:effort_metric}
\end{figure}

In Figure~\ref{fig:Tpar_Contraction_exp_1}, \ref{fig:Tpar_Gravity_exp_2}, and \ref{fig:Tpar_Rot_and_contract_exp_3}, we present the median parallel time among $5$ trials of each experiment executed with a particular load balancing algorithm. We employed the geometric partitioning algorithm from the Zoltan framework as well as our implementation of ``informed partitioning'', namely \textit{NoRCB} (non-orthogonal recursive coordinate bisection). Table~\ref{tab:lbcalls} summarizes the number of times the load balancer has been called during each experiment with every load balancing algorithm.

\begin{table}[h]
\begin{adjustbox}{center}
\begin{tabular}{lccc}
\toprule
                 & Gravity & Contraction & Rotation and Contraction \\
\midrule
NoRCB & $172$  &$183$ &$127$ \\
HSFC  & $1163$ &$1028$&$990$ \\
RCB   & $2038$ &$968$&$1154$\\
RIB   & $1820$ &$990$&$1043$ \\
\bottomrule
\end{tabular}
\end{adjustbox}
\caption{Summary of the number of load balancing calls required by each technique for each experiment.}
\label{tab:lbcalls}
\end{table}

In the Contraction experiment, we observe that ``informed partitioning'' outperforms the other techniques and decreases the wall-time by up to $76\%$. Therein, NoRCB can discover that the gas disk contracts and adapts the partition accordingly, thanks to the guidance of the average velocity vector. It is confirmed by our metric in Figure~\ref{fig:metric:contract}, which shows that the effort (i.e., the combined effect of the load balancing cost, the imbalance correction, and the load imbalance growth) induced by the partitioning is lower for informed partitioning. The slope of this curve is also less steep than the slope of the other geometric techniques. It indicates that the contraction of the gas (i.e., the density increase) has less effect on our technique. NoRCB has many advantages in this experiment. First, it allows an even partition of the interactions (although we use the algorithm on the particle positions). It is a consequence of the algorithm tessellation. Indeed, in this experiment, the closer we get to the center, the higher the density. Hence, by dividing the computational space radially  ``like a cake'' as shown in Figure~\ref{fig:contracting_gas_disks}, every processing element gets the same number of interactions, drastically decreasing the load imbalance (and its growth). Second, our implementation's high load imbalance correction capability, despite having a high cost due to a non-optimized parallel implementation, allows a low time per iteration post load balancing. Third, NoRCB triggers the load balancing mechanism only $172$ times, whereas the others use it more than seven times more ($1163$ for HSFC, $2038$ for RCB, and $1820$ for RIB), which again reduces the footprint of the load balancing mechanism. 

In the Gravity experiment, we notice that ``informed partitioning'' is less efficient than in the first experiment. However, it still reduces the application wall time by up to $15\%$. The reason for this lies in the change in the particle movement during application execution. Indeed, the gas starts by following a Brownian motion with a null average velocity but is slowly attracted to the ground by the gravitation force. After $3{,}000$ iterations, the particles are directed towards the ground. From this moment, NoRCB starts to shine because it can adapt the partitioning to the average velocity and effectively reduces the number of migrating particles. Hence, we see in Figure~\ref{fig:metric:grav} that ``informed partitioning'' (blue line) is not better than other techniques at the beginning (indistinguishable from other lines) but becomes effective only after $3{,}000$ iterations (the effort metric is lower for NoRCB at this point). Moreover, we report that NoRCB reduces the number of load balancing calls by up to $561\%$. It is partially due to the higher cost of load balancing induced by the non-optimized implementation of our technique.

\begin{figure}
    \centering
    \includegraphics[width=0.45\textwidth]{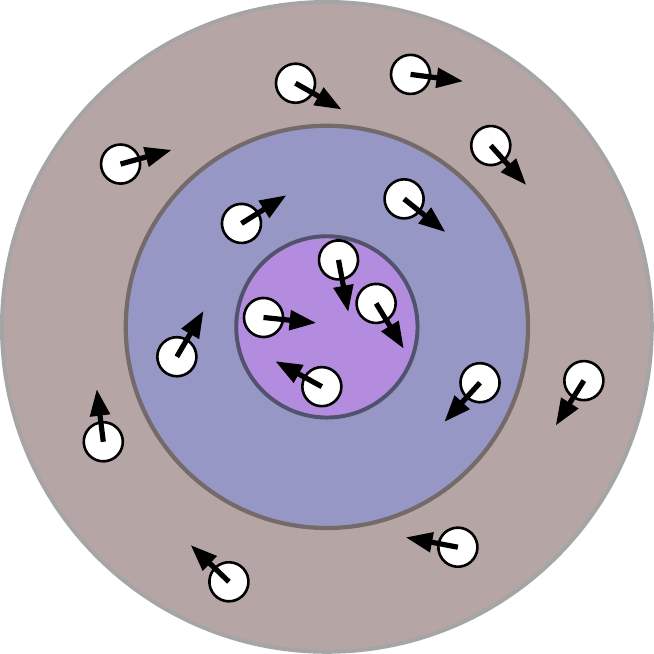}
    \caption{Example of circular partitioning that could be used in the rotation and contraction experiment. The colored areas describes the spaces assigned to the processing elements.}
    \label{fig:circlb}
\end{figure}

In the Rotation and Contraction experiment, NoRCB has no benefit over other geometric partitioning techniques. As we observe in Figure~\ref{fig:Tpar_Rot_and_contract_exp_3}, NoRCB is $3.2\%$ less efficient than HSFC, which performs the best. The reason is that the particles never have a straight movement due to the rotation of the gas disk and the attraction force. As mentioned earlier, when the particles follow a curved trajectory, the straight bisections prevent from reducing the load imbalance growth. Moreover, as shown in Figure~\ref{fig:metric:rotcontr}, HSFC seems to generate slightly less effort than the other techniques, which is confirmed in Figure~\ref{fig:Tpar_Rot_and_contract_exp_3}. In this experiment, a better partitioning should consider the angular velocity of the disk. Indeed, if the rotation is stronger than the contraction, a good partitioning would perform circular cuts as shown in Figure~\ref{fig:circlb}. It shows an example of circular partitioning where the space is divided into rings; in this figure each color denotes the space assigned to a processing element.

To conclude, we observed in this numerical study that using the velocity to drive the bisection in the recursive coordinate bisection algorithm allows an efficient reduction of the load imbalance growth, which improves the parallel application time by up to $76\%$ in the most suited experiment and $15\%$ in a harder one. Afterward, we saw that when the movement of the particles does not correspond to straight trajectories, this algorithm is not better than other geometric techniques. We also see that the effort metric presented in Section~\ref{sec:perf_metric} allows a better understanding of the reason why and, in particular \textit{when} load balancing algorithms perform well or not. Therefore, we highly suggest that researchers try to find ways to enhance existing load balancing techniques with informed partitioning. We also encourage the use of our effort metric to increase the understanding of the performance of load balancing algorithms during application execution.

\section{Discussion} \label{sec:discussion}

In a previous paper, we developed a load balancing paradigm in which we used information about the workload increase rate to ``underload'' the processing elements that are currently overloading, such that they will catch up due to their imbalance ``momentum''~\cite{Boulmier2019OnApplications}. This paradigm, combined with a simple stripe load balancing scheme, allowed us to improve the performance of a simulation of stochastic rock erosion application by up to $16\%$ compared to the classical HSFC algorithm from Zoltan. It appears that this previous work and the work presented in the present paper have similarities. In both works, the goal is to have long-lasting partitions by leveraging the dynamics of either the computational elements or the processing elements' workload itself. However, we use two distinct approaches to achieve the same objective.  \begin{figure}
    \centering
    \includegraphics[width=0.8\textwidth]{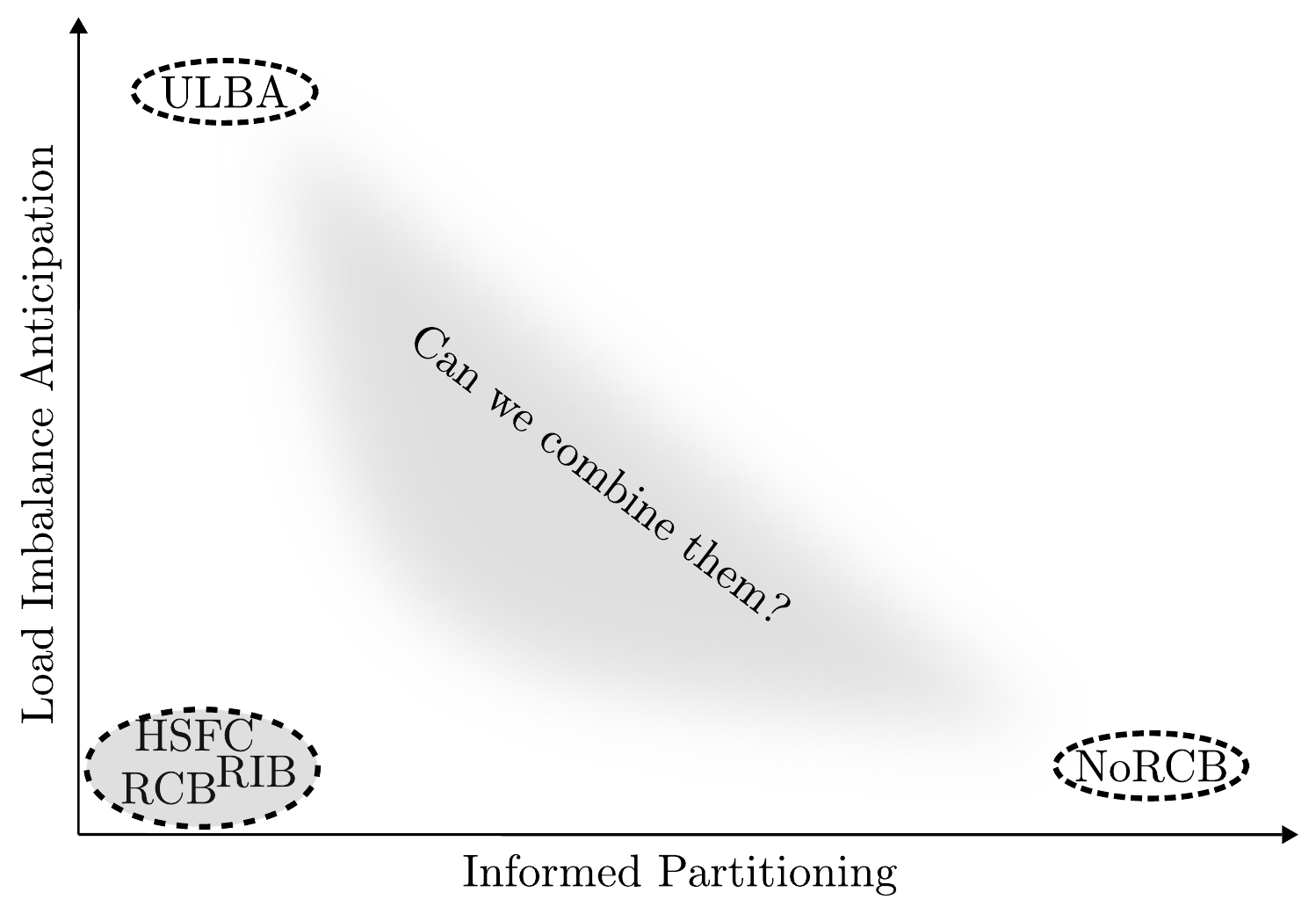}
    \caption{Load imbalance anticipation versus Informed Partitioning. Both methods tend to maximize the partition lifetime with different approaches. Is it possible to combine them to further improve load balancing techniques?}
    \label{fig:class_LAI_IP}
\end{figure} In Figure~\ref{fig:class_LAI_IP} we show a simple visualization of the two methods we have proposed. In the bottom left, classical load balancing methods does not use load imbalance anticipation nor informed partitioning to increase the partition lifetime. In the top left corner, ULBA~\cite{Boulmier2019OnApplications} anticipates the imbalance but does not adapt the partitions to it. Finally, in the bottom right corner, NoRCB (current paper) adapts the partition to reduce the load imbalance growth (ideally to suppress it). It is clear that it is not possible to combine the two methods as the goal of one is to cancel the other. Indeed, the ultimate goal of ``informed partitioning'' is to suppress the load imbalance growth. Hence, it makes no sense to combine a perfect ``informed partitioning'' technique with the load imbalance anticipation. In other words, the objective of these two techniques is orthogonal. However, we could imagine that, by accepting a little imbalance growth while using informed partitioning, we could employ anticipation to compensate. Such a combination may be more efficient than using a single method. Further research efforts are needed to confirm it, and this will be in the scope of future papers.

\section{Conclusion} \label{sec:conclusion}

In this paper, we presented the proof-of-concept of ``informed partitioning''. This concept employs application data to adjust the partitioning tessellation such that the load imbalance growth is reduced. Thus, in conjunction with the load balancing cost and the imbalance correcting capability, it improves the parallel application wall time. To illustrate the ``informed partitioning'' idea, we presented an algorithm that performs non-orthogonal recursive bisections in an N-Body simulation. In particular, this algorithm uses the velocity of the particles to guide the bisection axis, such that the number of particles that migrate between processing elements over time is reduced. We also proposed a new load balancing effort metric that incorporates all cost aspects of load balancing. This metric can be used to rank load balancing techniques by the amount of effort they cause to the application, and this at any time point during application execution, highlighting what technique is the most efficient at this stage. 

To assess the efficiency of ``informed partitioning'' and the use of our effort metric, we performed a numerical study on YALBB~\cite{XetqL/yalbb:Benchmark}, a home-made load balancing benchmark that solves the N-body problem with a short-range force. Therein, we compared the efficiency of ``informed partitioning'' with the geometric partitioning techniques available in the Zoltan~\cite{Devine2002} framework on a toy example and three concrete and more complex examples. Furthermore, we highlighted at what time during application execution our metric succeeded/failed to be efficient thanks to our novel effort metric. In particular, we observed that NoRCB performs the best when the particle trajectories are close to straight lines between two load balancing steps. In contrast, the benefit of our technique is reduced when their trajectory is strongly curved. In two out of three experiments, we observed performance improvement up to $76\%$ and $15\%$, while being a marginal $3\%$ slower than HSFC (best) in the last experiment. Finally, we pointed out that our effort metric does not reveal the source of the effort. In particular, it is difficult to understand if the effort comes from the load balancing cost, an inadequate imbalance correcting capability, or a significant load imbalance growth. 

To conclude, our plan regarding informed partitioning is many-fold. First, we plan to optimize our parallel implementation of NoRCB to test its efficiency on larger problems involving thousands of processing elements. Second, we will improve our effort metric to make it more explainable. Last, we plan to continue the research efforts on anticipation versus informed partitioning and how to combine them. 

\bibliographystyle{elsarticle-num.bst}
\bibliography{custom-ref}

\begin{thebibliography}{10}
\expandafter\ifx\csname url\endcsname\relax
  \def\url#1{\texttt{#1}}\fi
\expandafter\ifx\csname urlprefix\endcsname\relax\def\urlprefix{URL }\fi
\expandafter\ifx\csname href\endcsname\relax
  \def\href#1#2{#2} \def\path#1{#1}\fi

\bibitem{Garey1979}
M.~R. Garey, D.~S. Johnson, {Computers and Intractability: A Guide to the
  Theory of NP-Completeness}, W.H. Freeman {\&} Co., New York, USA, 1979.

\bibitem{Pearce2014}
O.~T. Pearce, M.~L. Adams, B.~R. De~Supinski, L.~Rauchwerger, V.~E. Taylor,
  {Load Balancing Scientific Applications: A Dissertation}, Ph.D. thesis, Texas
  A{\&}M University (2014).

\bibitem{Simon1997HowBisection}
H.~D. Simon, S.~H. Teng, {How Good is Recursive Bisection?}, SIAM Journal of
  Scientific Computing 18~(5) (1997) 1436--1445.
\newblock \href {https://doi.org/10.1137/S1064827593255135}
  {\path{doi:10.1137/S1064827593255135}}.

\bibitem{Pilkington1996DynamicCurves}
J.~R. Pilkington, S.~B. Baden, {Dynamic Partitioning of Non-Uniform Structured
  Workloads with Space Filling Curves}, IEEE Transactions on Parallel and
  Distributed Systems 7~(3) (1996) 288--300.
\newblock \href {https://doi.org/10.1109/71.491582}
  {\path{doi:10.1109/71.491582}}.

\bibitem{VanDriessche1995AnBalancing}
R.~Van~Driessche, D.~Roose, {An Improved Spectral Bisection Algorithm and its
  Application to Dynamic Load Balancing}, Parallel Computing 21~(1) (1995)
  29--48.
\newblock \href {https://doi.org/10.1016/0167-8191(94)00059-J}
  {\path{doi:10.1016/0167-8191(94)00059-J}}.

\bibitem{Karypis1998AGraphs}
G.~Karypis, V.~Kumar, {A Fast and High Quality Multilevel Scheme for
  Partitioning Irregular Graphs}, SIAM Journal of Scientific Computing 20~(1)
  (1998) 359--392.
\newblock \href {https://doi.org/10.1137/S1064827595287997}
  {\path{doi:10.1137/S1064827595287997}}.

\bibitem{Zhakhovskii2005ASimulation}
V.~Zhakhovskii, K.~Nishihara, Y.~Fukuda, S.~Shimojo, T.~Akiyama, S.~Miyanaga,
  H.~Sone, H.~Kobayashi, E.~Ito, Y.~Seo, M.~Tamura, Y.~Ueshima, {A New
  Dynamical Domain Decomposition Method for Parallel Molecular Dynamics
  Simulation}, in: IEEE International Symposium on Cluster Computing and the
  Grid (CCGrid), Vol.~2, 2005, pp. 848--854.
\newblock \href {https://doi.org/10.1109/CCGRID.2005.1558650}
  {\path{doi:10.1109/CCGRID.2005.1558650}}.

\bibitem{Fattebert2012}
J.-L. Fattebert, D.~F. Richards, J.~N. Glosli, {Dynamic Load Balancing
  Algorithm for Molecular Dynamics Based on Voronoi Clls Domain
  Decompositions}, Computer Physics Communications 183 (2012) 2608--2615.
\newblock \href {https://doi.org/10.1016/j.cpc.2012.07.013}
  {\path{doi:10.1016/j.cpc.2012.07.013}}.

\bibitem{Egorova2019ParallelSubdomains}
M.~S. Egorova, S.~A. Dyachkov, A.~N. Parshikov, V.~V. Zhakhovsky, {Parallel SPH
  Modeling using Dynamic Domain Decomposition and Load Balancing Displacement
  of Voronoi Subdomains}, Computer Physics Communications 234 (2019) 112--125.
\newblock \href {https://doi.org/10.1016/j.cpc.2018.07.019}
  {\path{doi:10.1016/j.cpc.2018.07.019}}.

\bibitem{Begau2015}
C.~Begau, G.~Sutmann, {Adaptive Dynamic Load-Balancing with Irregular Domain
  Decomposition for Particle Simulations}, Computer Physics Communications 190
  (2015) 51--61.
\newblock \href {https://doi.org/10.1016/j.cpc.2015.01.009}
  {\path{doi:10.1016/j.cpc.2015.01.009}}.

\bibitem{Deveci2016}
M.~Deveci, S.~Rajamanickam, K.~D. Devine, U.~V. Catalyurek, {Multi-Jagged: A
  Scalable Parallel Spatial Partitioning Algorithm}, IEEE Transactions on
  Parallel and Distributed Systems 27~(3) (2016) 803--817.
\newblock \href {https://doi.org/10.1109/TPDS.2015.2412545}
  {\path{doi:10.1109/TPDS.2015.2412545}}.

\bibitem{Boman2012Zoltan2}
E.~Boman, K.~Devine, V.~Leung, S.~Rajamanickam, L.~A. Riesen, M.~Deveci, U.~V.
  Catalyurek, {Zoltan2: Next-Generation Combinatorial Toolkit.}, Albuquerque,
  NM, USA, 2012.

\bibitem{Hirschmann2016}
S.~Hirschmann, D.~Pfluger, C.~W. Glass, {Towards Understanding Optimal
  Load-Balancing of Heterogeneous Short-Range Molecular Dynamics}, in: 23rd
  International Conference on High Performance Computing Workshops (HiPCW),
  IEEE, 2016, pp. 130--141.
\newblock \href {https://doi.org/10.1109/HiPCW.2016.027}
  {\path{doi:10.1109/HiPCW.2016.027}}.

\bibitem{Boulmier2021OptimalCriteria}
A.~Boulmier, N.~Abdennadher, B.~Chopard,
  \href{http://arxiv.org/abs/2104.01688}{{Optimal Load Balancing and Assessment
  of Existing Load Balancing Criteria}} (4 2021).
\newline\urlprefix\url{http://arxiv.org/abs/2104.01688}

\bibitem{Lieber2018HighlyModeling}
M.~Lieber, W.~E. Nagel, {Highly Scalable SFC-Based Dynamic Load Balancing and
  its Application to Atmospheric Modeling}, Future Generation Computer Systems
  82 (2018) 575--590.
\newblock \href {https://doi.org/10.1016/j.future.2017.04.042}
  {\path{doi:10.1016/j.future.2017.04.042}}.

\bibitem{H.MenonandN.JainandG.ZhengandL.Kale2012}
{H. Menon and N. Jain and G. Zheng and L. Kal{\'{e}}}, {Automated Load
  Balancing Invocation Based on Application Characteristics}, in: 2012 IEEE
  International Conference on Cluster Computing, 2012, pp. 373--381.
\newblock \href {https://doi.org/10.1109/CLUSTER.2012.61}
  {\path{doi:10.1109/CLUSTER.2012.61}}.

\bibitem{Procassini2004LoadCalculations}
R.~J. Procassini, M.~J. O'brien, J.~M. Taylor, {Load Balancing Of Parallel
  Monte Carlo Transport Calculations}, Tech. rep. (2004).

\bibitem{Zhai2018}
K.~Zhai, T.~Banerjee, D.~Zwick, J.~Hackl, S.~Ranka, {Dynamic Load Balancing for
  Compressible Multiphase Turbulence}, in: Proceedings of the 2018
  International Conference on Supercomputing - ICS '18, ACM Press, New York,
  New York, USA, 2018, pp. 318--327.
\newblock \href {https://doi.org/10.1145/3205289.3205304}
  {\path{doi:10.1145/3205289.3205304}}.

\bibitem{MarquezClaudio2013AApplications}
{M{\'{a}}rquez Claudio}, {Eduardo C{\'{e}}sar}, {Joan Sorribes}, {A Load
  Balancing Schema for Agent-Based SPMD Applications}, Proceedings of the
  International Conference on Parallel and Distributed Processing Techniques
  and Applications (PDPTA) (2013).

\bibitem{Hoare1961AlgorithmFind}
C.~A. Hoare, {Algorithm 65: Find}, Communications of the ACM 4~(7) (1961)
  321--322.
\newblock \href {https://doi.org/10.1145/366622.366647}
  {\path{doi:10.1145/366622.366647}}.

\bibitem{Devine2002}
K.~Devine, E.~Boman, R.~Heaphy, B.~Hendrickson, C.~Vaughan, {Zoltan Data
  Management Services for Parallel Dynamic Applications}, Computing in Science
  and Engineering 4~(2) (2002) 90--97.

\bibitem{SiebertScalableAndEfficient2014}
C.~Siebert, {Scalable and Efficient Parallel Selection}, in: Lecture Notes in
  Computer Science, Vol. 8384 LNCS, Springer, 2014, pp. 202--213.
\newblock \href {https://doi.org/10.1007/978-3-642-55224-3{\_}20}
  {\path{doi:10.1007/978-3-642-55224-3{\_}20}}.

\bibitem{DeRose2007DetectingSystems}
L.~DeRose, B.~Homer, D.~Johnson, {Detecting Application Load Imbalance on High
  End Massively Parallel Systems}, Springer, Berlin, Heidelberg, 2007, pp.
  150--159.
\newblock \href {https://doi.org/10.1007/978-3-540-74466-5{\_}17}
  {\path{doi:10.1007/978-3-540-74466-5{\_}17}}.

\bibitem{LaxmikantV.Kale2002}
{Laxmikant V. Kal{\'{e}}}, {The Virtualization Model of Parallel Programming:
  Run-time Optimizations and the State of Art}, in: LACSI, Albuquerque, 2002.

\bibitem{XetqL/yalbb:Benchmark}
\href{https://github.com/xetqL/yalbb}{{xetqL/yalbb: Yet Another Load Balancing
  Benchmark}}.
\newline\urlprefix\url{https://github.com/xetqL/yalbb}

\bibitem{Boulmier2019OnApplications}
A.~Boulmier, F.~Raynaud, N.~Abdennadher, B.~Chopard, {On the Benefits of
  Anticipating Load Imbalance for Performance Optimization of Parallel
  Applications}, in: International Conference on Cluster Computing (CLUSTER),
  IEEE, Albuquerque, NM, USA, 2019.
\newblock \href {https://doi.org/10.1109/CLUSTER.2019.8890998}
  {\path{doi:10.1109/CLUSTER.2019.8890998}}.

\end{thebibliography}

\end{document}